\documentclass[preprint]{elsarticle}
\usepackage{amssymb}
\usepackage{graphicx}
\usepackage{dcolumn}
\usepackage{bm}
\usepackage{fixltx2e}

\journal{Elsevier}

\begin{document}

\title{First principle investigations of the structural, electronic and magnetic properties of the new zirconium based full-Heusler compounds,  Zr\textsubscript{2}MnZ (Z = Al, Ga and In)}
\author[]{A. Birsan\footnote{A. Birsan $anca_{-}birsan@infim.ro$}$^{a,b}$}
\address{$^a$National Institute of Materials Physics, 105 bis Atomistilor Street, PO Box MG.7, 077125 Magurele-Ilfov, Romania. \\
 $^b$ University of Bucharest, Faculty of Physics, 105 Atomistilor Street, PO Box MG-11, 077125, Magurele-Ilfov, Romania.}

\begin{abstract}
The crystal structure, electronic and magnetic properties of the new full-Heusler compounds Zr\textsubscript{2}MnZ (Z=Al, Ga, In), were studied within the Density  Functional Theory (DFT) framework. The materials exhibit unique properties that connect the spin gapless semiconducting character with the completely compensated ferrimagnetism. In magnetic configurations, Zr\textsubscript{2}MnZ (Z=Al, Ga, In) crystallize in inverse Heusler structure, are stable against decomposition and have zero magnetic moment per formula unit properties, in agreement with Slater-Pauling rule. The Zr\textsubscript{2}MnAl compound presents spin gapless semiconducting properties with a energy band gap of 0.41 eV in the majority spin channel and a zero band gap in the minority spin channel. By substituting Ga or In elements, for Al in Zr\textsubscript{2}MnAl, semiconducting  pseudo band gaps are formed in the majority spin channels due to the different neighborhood around the manganese atoms, which decreases the energy of Mn's triple degenerated anti-bonding states.

\end{abstract}


\maketitle

\section{Introduction}
The technological progress in spintronics \cite{Uher1999}, solar cells \cite{Kieven2010} or thermoelectric devices \cite{Sakurada2005} requires the development of novel materials with designed desirable properties like half-metallic ferromagnetism (HMF) \cite{KublerPRB83,Wurmehl2006,Yin2015,Yin2013}, supraconductivity \cite{Winterlik2009} or non-magnetic semiconducting features \cite{Pierre1997}. 

The discovery of half-metallic properties \cite{Groot1983,Kubler1983} was the starting point for extensive investigations, to gain a profound understanding on the exceptional electronic structure of these high spin polarized materials. The half-metallic ferromagnetism would be distinguished when compounds show metallic as well as semiconducting properties at the same time. Formally, the materials with this peculiar property behave like metals for one spin channel and similar with semiconductors for the other spin channel. In this case, the electron-type carriers dominate the transport. 

The half-metallic ferrimagnets have advantages over the half-metallic ferromagnets because of the weak stray field and low magnetic moment achieved as result of the internal spin compensation \cite{Pickett,Park2002,KervanCAP2013,Min,BirsanInterm2013}. Moreover, if the ferrimagnetism resulted from the interaction occurred in the different sublattices is perfectly compensated with a net spin $m_{tot} = 0\:\mu_{B}$ the Completely Compensated Ferrimagnets (CCFs)\cite{Groot1991} are obtained also called Half-Metallic Completely Compensated Ferrimagnets (HM-CCFs) \cite{Wurmehl}. Thus, a complete spin polarization of carriers, theoretically should result when the spin-orbit interactions are nonexistent, which corresponds to the limiting case of zero temperature.  An interesting application for such materials may be as tips for Scanning Tunneling Microscopes (STM) because, in a soft magnetic material, the domain structure would not be distorted if the total magnetic moment is small. Apart from this, HM-CCFs are intensively studied to develop new compounds for stable spin-polarized electrodes in junction devices or integrated Spin-Transfer Torque Nano Oscillator (STTNO) for telecommunication. The half metallic completely compensated  ferrimagnetism properties were observed in double-perovskite structure oxides \cite{Pickett,Fuh2013}, thiospinels \cite{Park2001} or Heusler compounds \cite{Wurmehl2006,GalanakisPRB2007,GalanakisAPL2011,GalanakisJPhys2014,Balke,Graf2011,Odashima}.

In general, the electron-type carriers dominate the transport in the half metallic ferrimagnets, however holes can be also spin polarized, hence, an inverted band structure is obtained. In this particular case of half-metallicity, the materials act like topological insulators and high Curie temperature coexists with high resistance. The compounds with this peculiar transport property are known as spin gapless semiconductors because exhibit a semiconducting character in one spin channel and zero band gap in the other. 
Theoretical description of electronic structures and ground-state magnetism of several spin gapless semiconductors were provided via Density Functional Theory (DFT)\cite{Jia2014}. Some compounds like Mn\textsubscript{2}CoAl, have been grown as bulk samples or synthesized as thin films \cite{Ouardi2013,Jamer2013}. The experimental results confirm the stable phase with the crystalline structure with Hg\textsubscript{2}CuTi -prototype reported via DFT, and the spin gapless semiconducting properties. In this context, SGSs are being considered for spintronic applications, in general, and as promising substitutes for diluted magnetic semiconductors.

Cubic Heusler compounds, discovered in 1903 \cite{Heusler1903}, received recently intense research interest, because can exhibit high Curie temperature beside half-metallic properties at the Fermi level. Conventional full Heusler compounds, with the stoichiometric composition X\textsubscript{2}YZ  have the cubic L2\textsubscript{1} structure (space group Fm-3m and Cu\textsubscript{2}MnAl-prototype). However, when X atom is more electronegative than Y, the ordered inverse Heusler structure of X\textsubscript{2}YZ alloys  is obtained, containing four formula units per cubic unit cell, with X in 4a and 4c sites, Y in 4b sites, and Z in 4d sites (space group F\={4}3m and Hg\textsubscript{2}CuTi-prototype, also named inverse Heusler structure.).

Several Heusler compounds are known as ternary or quaternary half-metallic ferromagnets \cite{Wei2015,Wei2012,Xiong2014,Gao2015,BirsanCAP2014,BirsanJMMM2015}, half-metallic completely compensated ferrimagnets 
 \cite{Wurmehl,GalanakisPRB2007,JiaJMMM2014} or as spin gapless semiconductors 
  \cite{Ouardi2013,Wang2008,BirsanJMMM2013,Bainsla2015}. 
A combination of spin gapless semiconducting properties with completely compensated ferrimagnetism, with an integer $0\:\mu_{B}$ total magnetic moment per formula unit,  lead to Spin Gapless Completely Compensated Ferimagnetism (SG-CCF), in the class of Heusler compounds \cite{Fang2014}.

The possibility to obtain spin gapless semiconducting properties combined with zero magnetic moent in the ground state in zirconium based full-Heusler  Zr\textsubscript{2}MnZ (Z = Al, Ga and In) is analyzed in this paper, through theoretical description of electronic structure and ground-state magnetism via the Density Functional Theory. Zirconium based compounds were selected because exhibit low toxicity and being therefore  susceptible of convenient preparation and processing.

\section{METHOD OF CALCULATION}
The electronic structure and magnetic properties of Zr\textsubscript{2}MnZ (Z=Al, Ga, In) were theoretically studied based on first principle investigations via Density Functional Theory, with self-consistent Full Potential Linearized Augmented Plane Wave (FPLAPW) method, as implemented in WIEN2k code.\cite{Wien}. For the exchange and correlation interaction was used the Perdew Burke Ernzerhof (PBE) \cite{Perdew1996A, Perdew1996B} with Generalized Gradient Approximation (GGA). The muffin-tin model used for calculations, divides the unit cell into non-overlapping atomic spheres and interstitial regions. The muffin-tin radii (R\textsubscript{MT}) were selected as  2.35 a.u. for Zr, 2.40 a.u. for Mn and 2.18, 2.22 and 2.44 a.u. for Al, Ga and In, respectively. The cut-off condition for the number of plane wave selected to model the interstitial region was K\textsubscript{max}R\textsubscript{MT}=8 (where the K\textsubscript{max} represents the largest reciprocal lattice vector used for plane wave expansion and R\textsubscript{MT}the smallest muffin-tin radii).  A 27x27x27 mesh, containing 560 irreducible k points, was selected for Brillouin Zone integration, within the modified tetrahedron method \cite{Blochl1994}. The energy threshold  of -6 Ry was set between the core and valence states. The integrated charge difference between two successive iteration was less than $10^{-4}e/a.u.^{3}$ while the  convergence criterion for energy was $10^{-5}$ eV.

\section{RESULTS  AND DISCUSSIONS}
The  primitive cell considered for Zr\textsubscript{2}MnZ (Z=Al, Ga, In) compounds is the cubic face centered Hg\textsubscript{2}CuTi prototype structure, which is consistent with other X\textsubscript{2}YZ Heusler compounds studied experimentally in literature \cite{Ouardi2013}, having X atoms more electropositive than Y \cite{Kandpal2007}. In this structure, Zr atoms are located in two Wyckoff positions, having different neighborhoods 4a(0 0 0) and 4c(0.25 0.25 0.25) while the Mn and the Z atoms are located at 4b (0.5, 0.5, 0.5) and 4d(0.75 0.75 0.75), respectively.

The structural optimizations, via first-principle investigations were performed  for these new full-Heusler compounds Zr\textsubscript{2}MnZ (Z=Al, Ga, In), with  non-magnetic and  two antiferromagnetic configurations. In the first antiferrimagnetic configuration, all atoms had parallel magnetic spin orientation, while the second antiferromagnetic configuration, the atoms of the same type, Zr(4a)-Zr(4c) and Mn-Mn were anti-parallel coupled. The minimum values of the total energy for the first antiferromagnetic state were the lowest, comparing with those of non-magnetic and second antiferromagnetic cases. Therefore, the first antiferromagnetic  configurations were  used for further investigations of Zr\textsubscript{2}MnZ (Z=Al, Ga, In) compounds. From volume optimizations, the optimized lattice parameters were obtained for the first antiferromagnetic configurations of Zr\textsubscript{2}MnZ (Z=Al, Ga, In): 6.56 \.{A}, 6.58 \.{A} and 6.69 \.{A}, respectively.

The possibility to synthesize the new full-Heusler compounds Zr\textsubscript{2}MnZ (Z=Al, Ga, In), with inverse Heusler crystal structure, with at optimized lattice parameter, theoretically obtained, was analyzed by calculating the enthalpies of formation. The sum of the minimum values of the total energies calculated for Zr, Mn and Z(Z=Al, Ga, In), were subtracted from the minimum values of the total energy of Zr\textsubscript{2}MnZ (Z=Al, Ga, In) compounds for the first antiferromagnetic configurations. The negative values resulted for all compounds, lead to the conclusion that these alloys may be successfully prepared. 

\begin{figure}
 \begin{center}
    \includegraphics[scale=0.9]{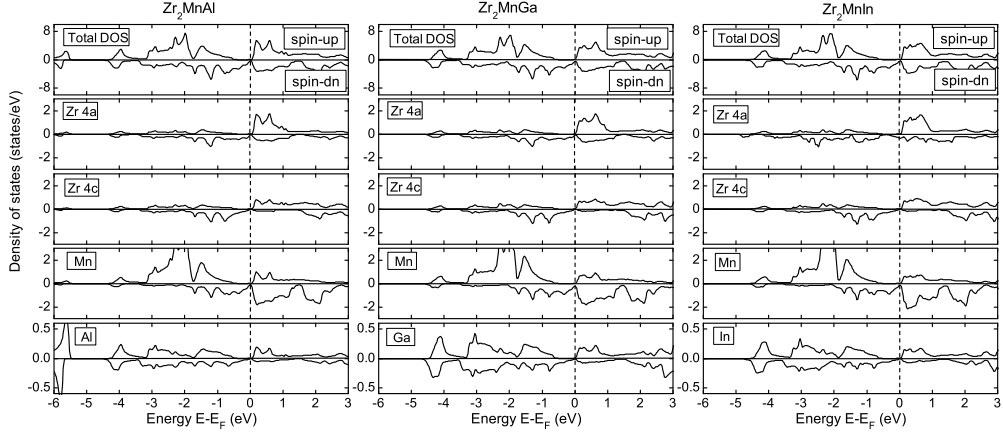} 
      \end{center}
   \caption{The site projected total and partial densities of states of Zr\textsubscript{2}MnZ (Z=Al,Ga,In) alloys performed at optimized lattice parameters.}
    \label{fig:Dos-all-Zr2Mn}
\end{figure}

The total and partial densities of states calculated at equilibrum lattice parameters are illustrated in Figure \ref{fig:Dos-all-Zr2Mn}. For Zr\textsubscript{2}MnAl, the energy gap from the majority spin  channel (spin up) lies a few eV below to the minimum edge of conduction band. In the minority spin channel (spin down), the Fermi level falls in a band gap often called zero or closed band gap. Therefore, the Zr\textsubscript{2}MnAl compound presents the property typical of spin gapless semiconductors and may allow a tunable spin transport. In the case of Zr\textsubscript{2}MnZ (Z=Ga, In), the spin up channels exhibit pseudo band gaps, due to cross intersection of Fermi levels, with densities of states provided by the anti-bonding d\textsubscript{t2g} states of manganese, having the lowest energy from conduction bands. In the spin down channels zero band gaps are present in both compounds, similar to the one formed in the Zr\textsubscript{2}MnAl material. For all compounds, in both spin channels, the significant contribution to density of states between -4.5 eV and -1.5 eV,  comes from d electrons of Mn and p electrons of main elements, located in Z position. Above the Fermi level, the states come from d electrons of Mn atoms in Zr\textsubscript{2}MnAl and d electrons of Mn and Zr(4a) for Zr\textsubscript{2}MnZ (Z=Ga, In). 

\begin{figure}
 \begin{center}
    \includegraphics[scale=0.9]{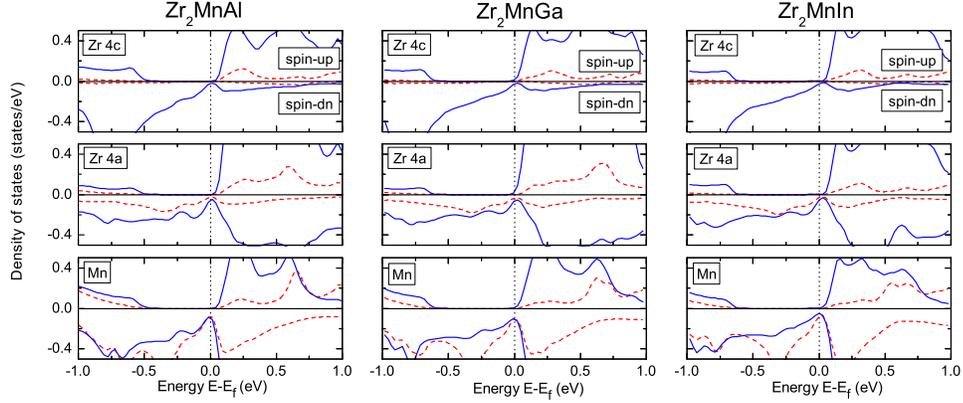} 
      \end{center}
   \caption{The densities  of states of double and triple degenerated states of Zr and Mn atoms, around the Fermi level, calculated at optimized lattice parameters for Zr\textsubscript{2}MnZ (Z=Al,Ga,In). The Fermi level, d\textsubscript{eg} and d\textsubscript{t2g} are illustrated with black dotted, red dashed  and blue solid line, respectively. }
    \label{fig:siteDOS-Zr2Mn}
\end{figure}

Figure \ref{fig:siteDOS-Zr2Mn} presents for all compounds, the contribution of double and triple degenerated states (d\textsubscript{eg} and d\textsubscript{t2g}, respectively) of Zr and Mn atoms, calculated around the Fermi level, at optimised lattice parameters. In Zr\textsubscript{2}MnAl compound, the highest bonding states from valence band, below the E\textsubscript{F}, belong to triple degenerated states of manganese  d\textsubscript{t2g}, while the lowest anti-bonding states from conduction band, come from the triple degenerated states  d\textsubscript{t2g} of Zr and Mn. As result,the energy gap from spin-up channel results due to Zr-Mn hybridization. The pseudo band gaps from the spin up channels of Zr\textsubscript{2}MnZ (Z=Ga, In), are also formed as result of d\textsubscript{t2g} states' hybridization of Zr-Mn atoms, the anti-bonding triple degenerated states of Mn from conduction band, having lower energy than the Fermi energy. In the spin down channel, for all compounds, the Mn and Zr(4a) atoms provide the few states around the Fermi level, hence, these may be called zero band gaps. However the compounds are theoretically studied here at the ground state (zero temperature), therefore, for materials with similar stoichiometry and studied experimentally, the amount of these states may increase/decrease. The double degenerated states, d\textsubscript{eg} of all atoms, play an insignificant contribution to density of states in all compounds, around the Fermi level.

\begin{figure}
 \begin{center}
    \includegraphics[scale=0.5]{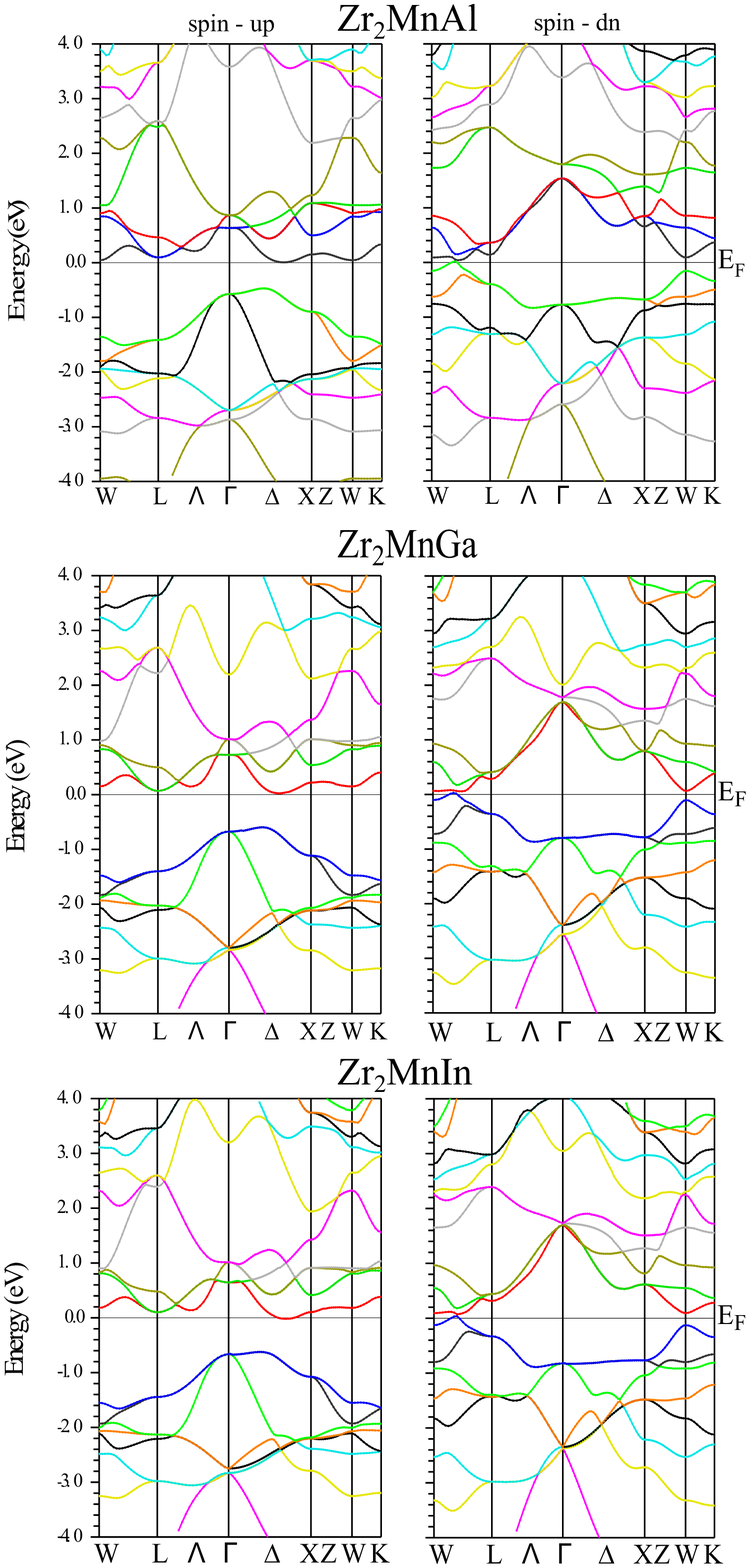} 
      \end{center}
   \caption{The band structure of Zr\textsubscript{2}MnZ (Z=Al,Ga,In) compounds for spin-up (left panels) and spin down channel (right panels), at equilibrium lattice parameter.}
    \label{fig:bands-Zr2Mn}
\end{figure}

The band structures calculated for optimized lattice parameters of Zr\textsubscript{2}MnZ (Z=Al,Ga,In) materials were illustrates in Figure \ref{fig:bands-Zr2Mn}. The Zr\textsubscript{2}MnAl alloy presents an indirect band gap of 0.41 eV, in the spin up channel with the higher bonding states from valence band located in the $\Delta$ point and the lowest anti-bonding states from the conduction band, distributed in the $\Delta$ and W  high symmetry points. In the case of Zr\textsubscript{2}MnZ (Z=Ga,In), the bonding states from valence bands located in the $\Delta$ point, determine with the triple degenerated anti-bonding states (d\textsubscript{t2g}) of Mn (having lower energy than Fermi energy), the pseudo band gaps from spin up channels. In the right pannel of the figure with the band structures of Zr\textsubscript{2}MnZ (Z=Al,Ga,In) materials, the zero band gaps may be observed, with the few states crossing the Fermi levels.

\begin{figure}
 \begin{center}
    \includegraphics[scale=0.7]{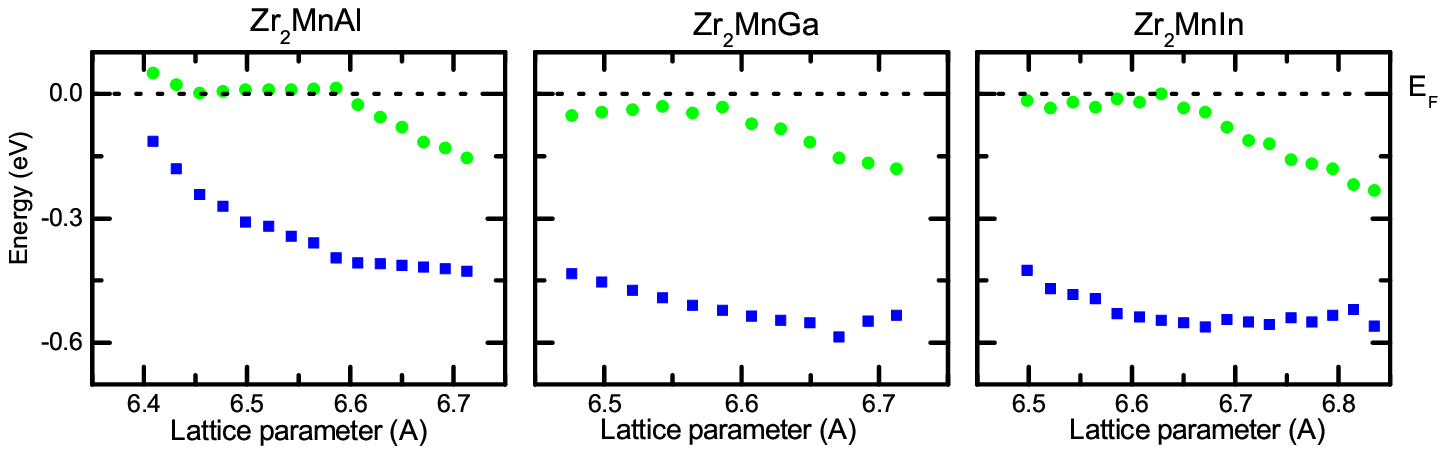} 
      \end{center}
   \caption{The positions of highest bonding states from the valence band ( blue solid squares) and of lowest anti-bonding states from conduction band ( green solid circles) of total DOSs for Zr\textsubscript{2}MnZ (Z=Al,Ga,In) as function of lattice parameters.}
    \label{fig:gapZr2MnZ}
\end{figure}

Figure \ref{fig:gapZr2MnZ} shows the change in the width of the energy band gaps from spin-up channels, for Zr\textsubscript{2}MnZ (Z=Al,Ga,In) materials as function of lattice parameters. It is obviously seen that the change in the lattice parameter affects the width of the energy band gaps, in all compounds. In Zr\textsubscript{2}MnAl alloy, the band gap increases by increasing of the lattice parameter. The largest value of the band gap's width is obtained for lattice parameter of 6.6 \.{A}, which corresponds to a volume optimization's increase of 2 $\%$. Above the lattice parameter of 6.6\.{A}, the spin gapless semiconducting properties of Zr\textsubscript{2}MnAl compound disappear, the Fermi level shifts into the conduction band and the width of the energy band gap from spin-up channel, decreases. In  Zr\textsubscript{2}MnZ (Z=Ga,In), above the optimized lattice parameters, the width of the pseudo band gaps formed in spin-up channels decreases with the increasing of the lattice constant and Fermi levels shift deep inside the conduction bands. Below the optimized lattice constant, the widths of energy pseudo band gaps have similar behavior as for Zr\textsubscript{2}MnAl compound. 

\begin{figure}
 \begin{center}
    \includegraphics[scale=0.7]{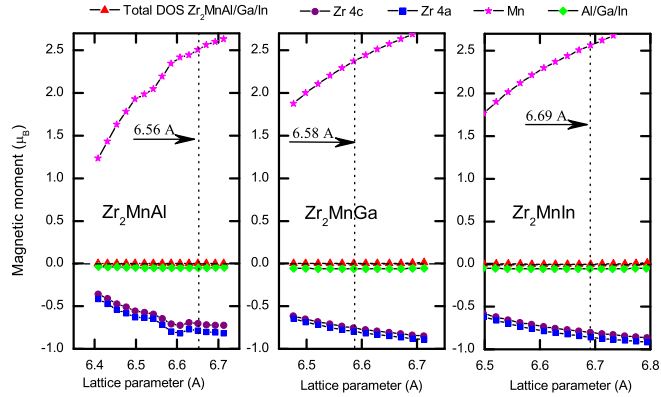} 
      \end{center}
   \caption{The total and partial magnetic moments for Zr\textsubscript{2}MnZ (Z=Al,Ga,In) compounds as function of lattice constants. }
    \label{fig:magnmom-Zr2Mn}
\end{figure}

All compounds studied exhibit perfectly compensated ferrimagnetism with calculated total magnetic moments of 0$\mu_{B}/f.u.$, which follow the Slater-Pauling rule for full-Heusler compounds with Hg\textsubscript{2}CuTi prototype crystalline structure (see Figure \ref{fig:magnmom-Zr2Mn}). Surprisingly, zirconium element which doesn't exhibit natively magnetic properties, shows magnetic behavior and a ferrimagnetic interaction occurs between the magnetic moments of Zr and Mn atoms. Moreover, zirconium atoms, located in different Wyckoff positions, are bound ferromagnetically. The magnetic moments of manganese, increase with the lattice parameter increase, in all compounds. The magnetic moments of zirconium atoms coupled ferromagnetically decrease with the lattice parameter increase and compensate the magnetic moment of Mn atoms. The calculated total and atomic spin magnetic moments and  optimized lattice parameters,  are given in Table 1. The main elements don't carry  significant magnetic moments. 

\begin{table}
\begin{center}
 \begin{tabular}{|c|c|c|c|c|c|c|}
 \hline  & a  & m\textsubscript{t} & m\textsubscript{Zr(4a)} & m\textsubscript{Zr(4c)} & m\textsubscript{Mn} & m\textsubscript{Z}   \\ 
 
   & (\.{A}) & ($\mu_{B}/f.u.$)& ($\mu_{B}/atom$) & ($\mu_{B}/atom$) & ($\mu_{B}/atom$)& ($\mu_{B}/atom$)   \\ 
 \hline Zr\textsubscript{2}MnAl & 6.56 & 0.00 & -0.77 & -0.69 & 2.44 & -0.04  \\
        Zr\textsubscript{2}MnGa & 6.58 & 0.00 & -0.79 & -0.75 & 2.37 & -0.06  \\ 
        Zr\textsubscript{2}MnGa & 6.69 & 0.00 & -0.85 & -0.80 & 2.57 & -0.05  \\ 
 \hline 
 \end{tabular}
 \end{center} 
 \caption{The optimised lattice parameters (in \.{A}), total spin magnetic moments (in $\mu_{B}/f.u.$), atomic spin magnetic moments ($\mu_{B}/atom$) calculated for the optimised lattice parameter, for Zr\textsubscript{2}MnZ (Z=Al, Ga, In). The differences between the calculated total magnetic moments and the  sum of atomic spin magnetic moments  represent the contributions of the interstitial regions.}
 \label{table:1}
 \end{table}

\section{Conclusions}
In conclusion, among the Zr\textsubscript{2}MnZ (Z=Al, Ga,In), the new Zr\textsubscript{2}MnAl compound exhibits a zero band gap on one spin channel and a semiconducting band gap in the other, with a completely compensated ferrimagnetism. The material follows the Slater-Pauling rule for a lattice parameter of the unit cell, up to 6.6 \.{A} and offers  the possibility to be used for spin-torque oscillators, spin-torque switched memory elements or for device applications in medicine. In addition the zero band gap from minority spin channel (spin down) provide an electronic structure extremely sensitive to external influences, because the electrons can move from occupied states to empty states without any threshold energy. The Zr\textsubscript{2}MnZ (Z=Ga,In) compounds present nearly spin gapless semiconducting properties, determined by the pseudo energy band gaps from majority spin channels. All compounds exhibit zero magnetic moment per formula unit, at optimized lattice parameter, in the ground state. 

\section{ACKNOWLEDGMENTS}
The author thanks Dr. P. Palade for his support, and Dr. V. Kuncser for helpful discussions. This work was financially supported by the Romanian National Authority for Scientific Research through the CORE-PN45N projects and by the strategic grant POSDRU/159/1.5/S/137750, cofinanced by the European Social Found within the Sectorial Operational Program Human Resources Development 2007-2013.

\bibliographystyle{elsarticle-harv}

\end{document}